\documentclass[12pt]{article}
\usepackage{amsmath, amsthm, amssymb}
\usepackage{natbib}
\usepackage{amssymb}
\usepackage{graphicx}
\usepackage{multirow}
\usepackage[normalem]{ulem}
\usepackage{color}

\date{}
\begin{document}

\title{The propagation of relativistic jets in external media}
\author{Omer Bromberg$^1$, Ehud Nakar$^2$, Tsvi Piran$^1$, Re'em Sari$^1$\\
\small $^1$ Racah Institute of Physics, The Hebrew University, 91904 Jerusalem, Israel\\
\small $^2$ The Raymond and Berverly Sackler School of Physics and Astronomy,\\
\small Tel Aviv University, 69978 Tel Aviv, Israel}
\maketitle

\begin{abstract}
Relativistic jets are ubiquitous in astrophysical systems that contain compact objects.
They transport large amounts
of energy
to large distances from the source, and their interaction with the ambient
medium has a crucial effect on the evolution of the system.
The propagation of the jet is characterized by the formation of a shocked "head"
at the front of the jet which dissipates the jet's energy and a cocoon
that surrounds the jet and potentially collimates it.
We present here a self consistent, analytic model that follows the evolution of the jet
and its cocoon, and describes their interaction.
We show that the critical parameter that determines the properties of the jet-cocoon system
is the dimensionless ratio between the jet's energy density and the rest-mass
energy density of the ambient medium.
This parameter, together with the jet's injection angle, also determines whether the jet is
collimated by the cocoon or not.
The model is applicable to relativistic, unmagnetized, jets on all scales and may be used to
determine the conditions in AGNs jets as well as in GRBs or microquasars.
It shows that AGN and microquasar jets are hydrodynamically collimated
due to the interaction with the ambient medium, while GRB
jets can be collimated only inside a star and become uncollimated once they breakout.
\end{abstract}

\section{Introduction}

Relativistic jets are observed in many astrophysical systems which
host compact objects, such as radio galaxies, gamma-ray bursts
(GRBs) and microquasars. These jets appear in a variety of lengths,
durations and energy scales, and propagate in different types of
media. However, despite the differences in their characteristics,
which are many orders of magnitudes apart, the interactions of the
jets with their surroundings lead to similar results, such as energy
injection into the ambient medium and its feedback on the jet.
Understanding these common
phenomena can provide valuable insights on puzzles such as what type
of medium can collimate the jet and how much energy is injected into
the surrounding matter. These in turn can be used to study the
heating of the interstellar and intergalactic medium by active
galactic nuclei (AGNs) and the fate of jets and stellar envelops of
collapsing massive stars during gamma-ray bursts.

The interaction of a jet with the external medium was studied
extensively in many different scales, using analytic and numerical
methods. Initially it was studied in the context of radio loud AGNs
\citep[][]{BlandRees74,Scheuer74}. They showed that the propagation
of the jet generates a double bow-shock structure at the head of the
jet. Energy and matter that enter this structure are pushed aside
due to a high pressure gradient and create a hot cocoon around the
jet. The cocoon, in turn, applies pressure on the jet and compresses
it. \citet{BegelCiof89}  calculated the head's velocity and the
cocoon expansion rate, assuming that the cocoon and the head are
supported by the ram-pressure of the ambient medium. Their
calculations assumes Newtonian head velocities and rely on the
knowledge of the cross-section of the head, which they estimated
phenomenologically using  the size of the radio lobes observed
around jets of radio loud AGNs.  \citet{MW01} calculated the Lorentz
factor of a relativistic head assuming a conical jet (i.e., no
collimation).
\citet{Matzner03} extended the model of
\citet{BegelCiof89} to include both Newtonian and relativistic head
velocities, assuming a conical jet as in \citet{MW01}. Later, \citet{LazzBeg05} used this
model to calculate the opening angle and the properties of the jet
at breakout from a surface of a collapsing star. They accounted
for the reduction in the opening angle of the jet due to collimation, but
ignored the dissipation of energy in shocks that form within the jet
as a result of such a collimation. This led to an incorrect dependence of the opening
angle on the cocoon's pressure. An attempt to include the
effects of these shocks was made by \citet{Morsony07}.

Apart from
the analytical studies the numerical approach was also given a lot of
attention over the years. Simulations were conducted in the context
of extra galactic jets \citep[e.g.][]{Marti95,Marti97,Aloy99,Hughes02} as well as
GRB jets that propagate inside a star
\citep[e.g.][]{Zhang03,Morsony07,Mizuta09}. All these works showed
the basic features discussed above, i.e. the formation of a jet
head and a hot cocoon. A collimation shock at the
base of the jet
is also evident in the simulations, whenever the jet is collimated by
the cocoon. However, despite the extensive numerical and analytical
efforts, to this date there is no simple analytic description of the
time evolution of the jet-cocoon system.
As a result, there is no quantitative understanding
of the collimation process, and its effect on the jet angle.

The goal of this work is to provide a simple self consistent,
analytic description, for the time evolution of an unmagnetized
relativistic jet and the cocoon that forms  around it, when it
propagates in a density profile that is suitable for most
astrophysical systems. The key point in our analysis is the
treatment of the collimation shock that forms at the base of the jet. This shock
is an inevitable consequence of collimation in a supersonic jet, as
it dissipates part of the jet's energy, generating the pressure
needed to counterbalance the cocoon's pressure. It is crucial for
the proper modeling of the system, since it sets the width of the jet
and controls its propagation velocity.
We base our solution on the earlier analysis of
\citet{BegelCiof89,Matzner03}, and incorporate the geometry of
the collimation shock as it is analyzed by \citet{KomisFall97, BL09}. We find that
the jet evolution can be of two types: collimated and un-collimated according to
the strength of the jet - cocoon interaction and the
collimation shock.
We quantify the transition between these regimes and derive models for the jet and the cocoon
evolution in each regime. We
test the validity of our approximations by comparing the
results from our model
with output of various numerical simulations.

The paper is constructed as follow:
In \S2 we present a general nontechnical description of the system,
its ingredients and the different stages in its evolution.
In \S3 we present our model and derive the criterion that distinguishes
between the two collimation regimes. We discuss the geometry of the collimation
shock in each regime and analyze how it affects the temporal evolution of
the system. We also provide approximated analytic solutions in each of the regimes.
We compare our analytical model to hydrodynamic simulations in \S4.
In \S5 we briefly consider jet collimation in various astrophysical
environments. Thorough study of specific systems
is lengthy and will be presented in following papers.
In \S6 we summarize the main results.
The full list of results, that specifies the behavior of the physical
quantities in the entire parameter space is summarized in table 1 and in
appendix B.

\section{A schematic description of the system}

Consider an un-magnetized, sufficiently fast\footnote{Defined later
at \S3.1, eq. (\ref{eq:fast}).} relativistic jet with a luminosity
$L_j$, injected at a fixed opening angle $\theta_0$, into the
surrounding medium. We assume that the
entire system (jet, cocoon and ambient medium) is axisymmetric
and use cylindrical coordinates ($z,r$). We further
approximate the ambient medium density, $\rho_a$, to depend only on
the hight, i.e. $\rho_a(z)$. This is a good approximation even  in
cases where $\rho_a$ is spherically symmetric (e.g., a stellar
envelope), since the opening angles of the jet and the cocoon are
small whenever the the ambient medium plays an important role and
collimates the jet.

\begin{figure}
  \includegraphics[width=6in]{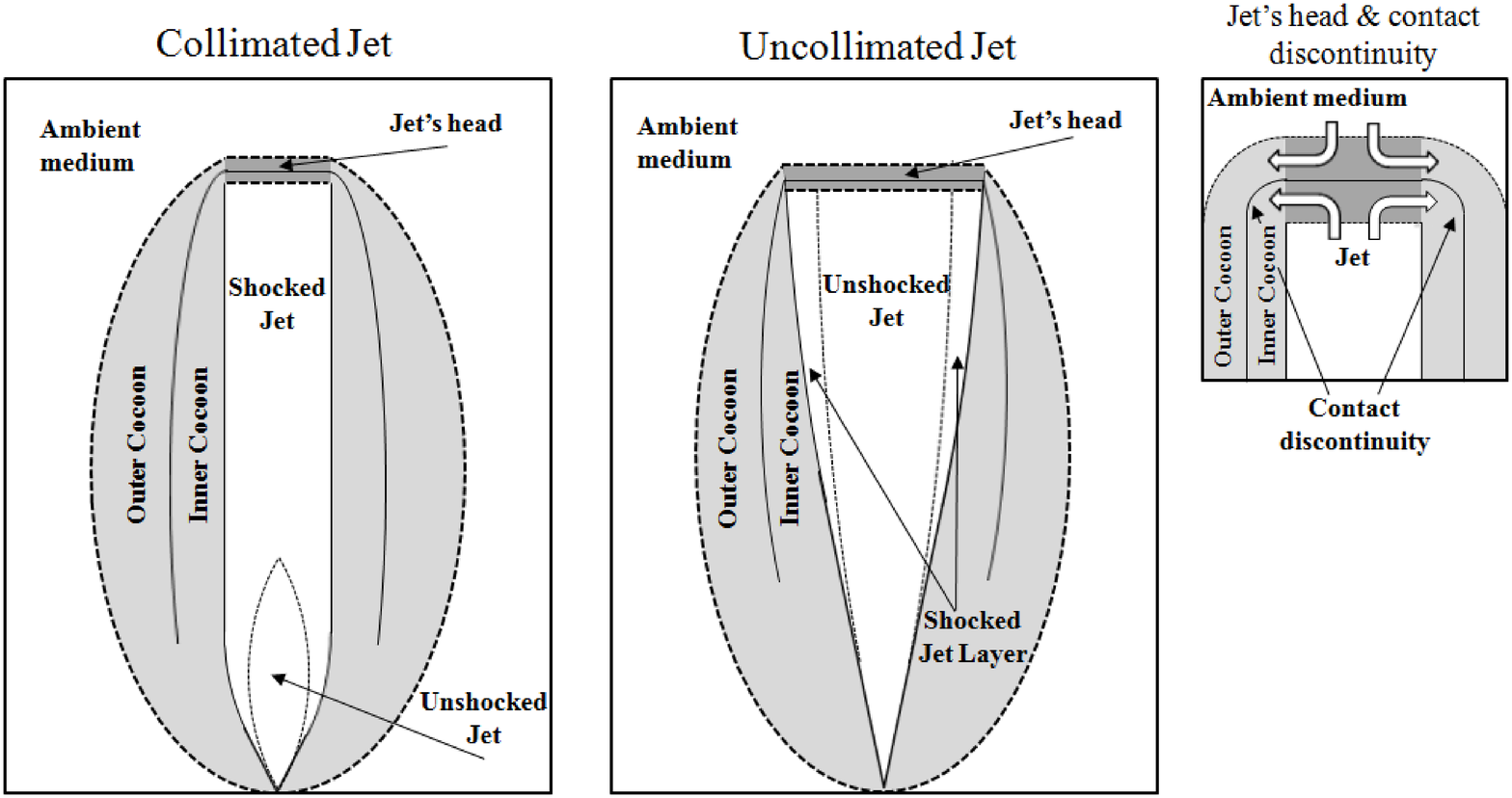}
  \caption{A schematic description of the jet geometry in the two
  collimation regimes.
  Left: A collimated jet; Center:  An un-collimated jet.
  In both panels the basic ingredients of the model are evident:
  the jet (divided into a shocked and and un-shocked part), the jet's head, the cocoon and the ambient medium.
  Also shown are the collimation shock and the contact discontinuity.
  The collimation shock splits the jet to an un-shocked region and a shocked region.
  The contact discontinuity separates the jet
  material that enters the head from the ambient medium.
  This discontinuity extends to the cocoon and divides it to an inner and an outer part.
  The cocoon expands into the ambient medium behind a shock that extends the
  forward shock at the head.
  All shocks are marked with dashed lines, and the contact discontinuities
  with solid lines.
  Right: A closeup of the jet's head and the contact discontinuity.
  The matter flows into the head through a forward and a reverse shock,
  and from there to the cocoon as
  illustrated by the four arrows.}\label{jet_fig}
\end{figure}

The jet propagates by pushing the matter in front of it, leading to
the formation at the jet's front of a forward shock and a reverse
shock that are separated by a contact discontinuity. We refer to
this structure as the jet's head. Matter that enters the head
through the shocks is heated and flows sideways  since its pressure
is higher than that of the surrounding matter (see fig. 1). This
leads to the formation of a pressured cocoon around the jet. A
contact discontinuity divides the cocoon into an inner, light, part
containing the jet material which crossed the reveres shock, and an outer part of the
heavier shocked medium. The pressure in both sides of the
discontinuity is equal, but since the plasma is much more tenuous
in the inner cocoon, the sound speed there is much larger than in the outer
cocoon. This allows a causal connection along the cocoon, leading to
pressure equilibration in the vertical direction and a more uniform
distribution of the energy (see further discussion in
\ref{sec:model}). Since our analysis needs only the cocoon's
pressure we disregard in the following the cocoon's inner structure.
If the cocoon's pressure is sufficiently high, it collimates the jet
and reduces its opening angle. This changes the jet's propagation velocity
and the energy flow into the cocoon. We classify two regimes of the
system: collimated and un-collimated, according to how strongly the
jet is affected by the cocoon's pressure. Fig. (\ref{jet_fig})
depicts a schematic description of the system in these two regimes.

In the collimated regime (fig. \ref{jet_fig}, right panel), the
cocoon's pressure  collimates the jet and reduces substantially its
opening angle. Since the jet is supersonic, the collimation leads to
the formation of an oblique shock at the base of the jet. This shock
deflects the jet flow-lines and generates a pressure that
counterbalances the cocoon's pressure. To maintain the required
pressure the shock curves toward the jet's axis, until it converges
at some altitude, above which the collimation is complete (see
\S3.1). The geometry of the collimation shock sets the jet's
cross-section at the head to be much smaller than the cross section
of an  un-collimated  jet. Consequently the jet applies a larger
ram-pressure on the head and pushes it to higher velocities. The
faster motion reduces the rate of energy flow into the cocoon. At
the same time the cocoon's height increases at a faster rate,
resulting in a larger volume and a decrease in the cocoon's
pressure.
There is a limit to the head velocity above which the
cocoon's pressure is too low to effectively collimate the jet.
We show that this  occurs when $L_j/(z^2\rho_a c^3)\simeq\theta_0^{5/3}$,
corresponding to a head's Lorentz factor
$\Gamma_h  \simeq \theta_0^{-1/3}$.
Therefore for typical initial opening angles $\theta_0>1^\circ$, the head velocities
in the collimated regime can be at most  mildly relativistic.

The un-collimated regime is characterized by larger values of $\Gamma_h$ and a
cocoon pressure which is insufficient to collimate the jet in an appreciable amount.
The jet remains conical to a good approximation and the collimation shock remains at the edges of the jet and
does not converge onto the jet's axis. This results in a coaxial jet structure composed of an inner fast spine surrounded by a denser layer of the shocked jet material, having a lower Lorentz factor (Fig. \ref{jet_fig}, left panel).
When $\Gamma_h>\theta_0^{-1}$, the head  moves so fast that
different parts of the jet's head become causally disconnected and
energy can flow into the cocoon only from a small region of the
head. This further reduces the  cocoon's pressure.  At even larger
Lorentz factors the reverse shock at the head becomes weak, and it
no longer affects the jet, which can be considered as propagating in
a vacuum. The forward shock continues,  however, to gather matter from
the ambient medium in front of the jet and to accelerates it to a
Lorentz factor similar to that of the jet. A small fraction of this
shocked matter continues to stream into the cocoon and feeds with relativistic
particles.

Following the above description we divide the system into five main
elements (see figure \ref{jet_fig}): un-shocked jet, shocked jet (by
the collimation shock), jet's head, cocoon (containing an inner
light part and an  outer heavy part) and ambient medium. The system
dynamics is determined by the following relations between these
region. (i) The jet's head velocity is set by balancing the
ram-pressure applied on the forward shock (by the ambient medium)
with the ram-pressure applied on the reverse shocks (by the shocked
or unshocked jet, depending on the collimation regime). The head
velocity determines the cocoon height
 and the energy injection rate
into it. The jet head has the highest pressure in the system. (ii)
The pressure in the cocoon is set by its size and by the energy
injected into it from the head. (iii) The velocity of the shock that
inflates the cocoon into the ambient medium is set by the balance
between  the ram-pressure of the ambient medium on that shock and
the cocoon pressure. (iv) The pressure in the shocked jet
is equal to the cocoon pressure. The collimation shock structure is
set to build up this pressure in the jet. This structure, in turn,
determines the jet head cross section and thus the ram pressure
applied on the reverse shock. (v) The un-shocked jet properties are
determined by the inner engine.


\section{The jet-cocoon model}\label{sec:model}

Our model contains the five elements discussed above, where the
ambient medium serves as a fixed background. Given a jet with a
luminosity $L_j$, an injection angle, $\theta_0$, and a medium density
profile, $\rho_a(z)$, we calculate the time dependent quantities:
the head velocity, $\beta_h$ (predominantly in the $z$ direction),
the cocoon's pressure, $P_c$, the
cocoon expansion velocity, $\beta_c$ (predominantly in the $r$ direction),
and the jet's cross-section
$\Sigma_j$. We use the subindices $j,h,c,a$ to designate quantities
related to the jet, the jet's head, the cocoon and the ambient
medium respectively. The distinction between the shocked and
un-shocked jet is relevant only in the collimated regime, and we use
it when we discuss this regime. The subindex $j$ is used to describe
general properties of the jet (like the jet dimensions) and when it
is unimportant which of the two region (shocked or unshocked jet) is
considered, e.g., the jet luminosity is equal in the two regions and
is denoted $L_j$.

We begin by discussing the main assumptions of the model.
We take $L_j$, $\theta_0$ and $\rho(z)$ as given, and derive
equations for $\beta_h$, $P_c$, $\beta_c$ and $\Sigma_j$.
The later differs between the two collimation regimes, and
therefore we present the relevant solution separately
in each regime.

The head velocity is set by balancing the ram-pressure applied on the forward and reverse shocks \citep{BegelCiof89,Matzner03}:
\begin{equation}\label{ram_head}
 \rho_j h_j \Gamma_{j}^2\Gamma_{h}^2\left(\beta_j-\beta_h\right)^2 + P_j = \rho_a h_a\Gamma_h^2\beta_h^2 + P_a,
\end{equation}
where $\rho$, $P$ and  $h=1+4P/\rho c^2$ are the mass density,
pressure and dimensionless specific enthalpy of each fluid, measured in the fluid's rest frame,
and $\Gamma\beta$ is the proper fluid velocity in the frame of the ambient medium.
When the temperature of the ambient medium is non-relativistic and as long as
the reverse shock is strong, $P_a$ and $P_j$ can be ignored.
Under these conditions the head's velocity is  \citep{Matzner03}:
\begin{equation}\label{bh}
\beta_h=\frac{\beta_j}{1+\tilde L^{-1/2}},
\end{equation}
where the dimensionless parameter
\begin{equation}\label{tild_L}
\tilde L\equiv\frac{\rho_j
h_j\Gamma_j^2}{\rho_a}\simeq\frac{L_j}{\Sigma_j\rho_ac^3}
\end{equation}
represents the ratio between the energy density of the jet
($L_j/\Sigma_j c$) and the rest-mass energy density of the
surrounding medium at the location of the head. Note that by
defining $\tilde L$ as such, we implicitly assume that the energy
lost in the jet due to work against the pressure of the cocoon is
negligible and that radiation losses are negligible as well. We
discuss this assumption at length in Appendix A, and
show that it is valid as long as the jet injection angle,
$\theta_0$, is not too large.
As we show in the following sections, $\tilde L$ is the
critical parameter that determines the evolution of the jet and the
cocoon, while the combination $L\theta_0^{4/3}$
distinguishes between collimated and un-collimated
jets. We stress that $\tilde L$ may vary with the propagation of the
jet, even if the jet luminosity is constant, depending on the
density profile of the ambient medium and on the behavior of the
jet's cross section. Thus, a jet can switch between the different
collimation regimes, from collimated to un-collimated and vice versa.
Specifically a collimated jet that encounters a sharp density
decline may becomes uncollimated.

In the limits $\tilde L\ll1$ or $\tilde L\gg1$, eq. (\ref{bh}) can be linearized \citep{Matzner03}:
When $\tilde L\ll1$
\setcounter{equation}{4}
\begin{equation}\label{bh_nr}
\tag{4a}
\beta_h\simeq\tilde L^{1/2},
\end{equation}
corresponding to a non-relativistic head velocity. If
$1\ll\tilde L\ll4\Gamma_j^4$
\begin{equation}\label{gammah_r}
\tag{4b}
\Gamma_h\simeq\sqrt{\frac{1}{2}}\tilde L^{1/4},
\end{equation}
which corresponds to a relativistic proper velocity of the head,
$\Gamma_h\beta_h>1$  \citep[e.g.][]{MW01}. When $\tilde
L\gg4\Gamma_j^4$, it follows from eq. (\ref{bh})  that
$\Gamma_h\simeq\Gamma_j$, implying a non-relativistic reverse shock and a
negligible energy flow of the shocked jet material into the cocoon.
This is equivalent to the requirement that the volumetric-enthalpy ratio of the
jet and the ambient medium, $\frac{\rho_jh_j}{\rho_ah_a}>\Gamma_j^2$, and
is similar to the criterion given by \citep{Sari95} for the
case of a cold jet.

The pressure in the cocoon is sustained by a continuous flow of
energy from the head. At any given moment the total energy in the
cocoon is $E_c=\eta L_j(t-z_h/c)$, where the second term in the
brackets represents the fraction of the energy which is carried by
the relativistic jet and hasn't reached the cocoon yet
\citep{LazzBeg05}. The parameter, $\eta$, varies  between 0 and 1.
It stands for the fraction of the energy that flows into the head and enters
later into the cocoon. We assume that all the energy in the region of the head that is in
causal connection with the cocoon flows into the cocoon, so $\eta$ is the fraction of that region
out of the total head volume.
When $\Gamma_h\gg1$ most of
the energy that flows out is initially tapped to the bulk motion of
the fluid and it can't contribute to the pressure. However, later,  as
the matter enters the cocoon it spreads sideways and decelerates
exponentially to $\Gamma\simeq1$, in a similar manner to a rapidly
spreading jet in a GRB afterglow \citep[e.g.][]{Granot00,Piran00}.
Therefore practically all the energy that flows into the cocoon is
available to generate pressure.

The energy in the cocoon is shared between two parts which are
separated by a contact discontinuity: an outer part made of the
shocked ambient medium with a typical density $\sim\rho_a$,
and an inner part that consists of the lighter jet material that crossed
the reverse shock and enters from the head (see fig. 1).
The outer cocoon is supported from the sides by the ram pressure of
the surrounding medium, which balances its
pressure. This results
in a lateral expansion velocity of $\beta_c=\sqrt{P_c/\rho_ac^2}$
which is just below the sound speed in the outer cocoon as long as
$\beta_c$ is sub relativistic. In the inner cocoon the sound speed
is relativistic and typically much faster than $\beta_c$. Thus,
matter on both sides of the contact discontinuity is in causal
contact in the lateral direction and any pressure difference in this
direction is smoothed out. In the longitude direction the inner cocoon connects
the upper parts of the cocoon with lower parts which otherwise would
have been out of causal contact. This suppresses the pressure drop
in the vertical direction, and in cases where the head is
sub relativistic leads to a uniform distribution of energy
in the cocoon. The size of the region connected by the inner cocoon is
determined by the survival length of the contact discontinuity,
which is set by the growth rate of Kelvin-Helmholtz instabilities.
The small shear-velocity differences on both sides of the contact
render the growth rate small and stabilize the contact
discontinuity.
Indeed numerical simulations show that the contact
discontinuity remains intact for a considerable fraction of the
cocoon's length, and that the pressure drop is suppressed (see \S4).
We stress that the uniform pressure approximation
should also hold rather well in cases
where $P_c$ develops some vertical gradient since
the conditions in the jet's head depends only weakly on the pressure
gradient.

To calculate the cocoon's pressure we make the following
approximations: First we take the cocoon's shape to be a cylinder
with a hight $z_h=\int\beta_hdt$ and a radius $r_c=\int\beta_cdt$.
Second, based on the discussion above, we approximate the
energy density to be uniformly distributed within the cocoon.
Third, over a wide range of parameters and especially in the systems
relevant for us, the pressure in the cocoon is radiation dominated,
thus we use  in our analysis an adiabatic index of $4/3$. Under
these approximations the cocoon's pressure satisfies:
\begin{equation}\label{Pc}
P_c=\frac{E}{3V_c}=\frac{\eta}{3\pi c^3}\frac{L_j\int(1-\beta_h)dt}{\int\beta_hdt \left(\int\beta_cdt\right)^2}.
\end{equation}
To calculate the transverse velocity of the cocoon we take the average density of the medium
$\bar{\rho}_a=1/V_c(z)\int\rho_a(z)dV$, where $V_c(z)$ is the
volume of the cocoon, and obtain the lateral expansion velocity:
\citep[e.g.][]{BegelCiof89}:
\begin{equation}\label{bc}
\beta_c = \sqrt{\frac{P_c}{\bar{\rho}_a(z_h)c^2}}.
\end{equation}
Approximating $\int\beta_hdt\sim\beta_ht$ and $\int\beta_cdt\sim\beta_ct$ and
substitute  into eq. (\ref{Pc}) we obtain the cocoon's pressure:
\begin{equation}\label{Pc_ann}
P_c=\Xi_a\left(\frac{L_j\rho_a}{3\pi c}\right)^{1/2}\tilde L^{-1/4}t^{-1},
\end{equation}
where $\Xi_a$ is a constant of order unity that depends on the
density profile of the ambient medium, and on the collimation regime.
The value of $\Xi_a$ in each regime is given in the appendix,
where we list the full time dependent solutions of eqn.
(\ref{bh}), (\ref{Pc}) and (\ref{bc}) in the limits: $\tilde L\ll1$ and $\tilde L\gg1$.

\subsection{The collimated regime ($\tilde L<\theta_0^{-4/3}$).}\label{sec.collim}

\begin{figure}[h]
  \includegraphics[width=5in]{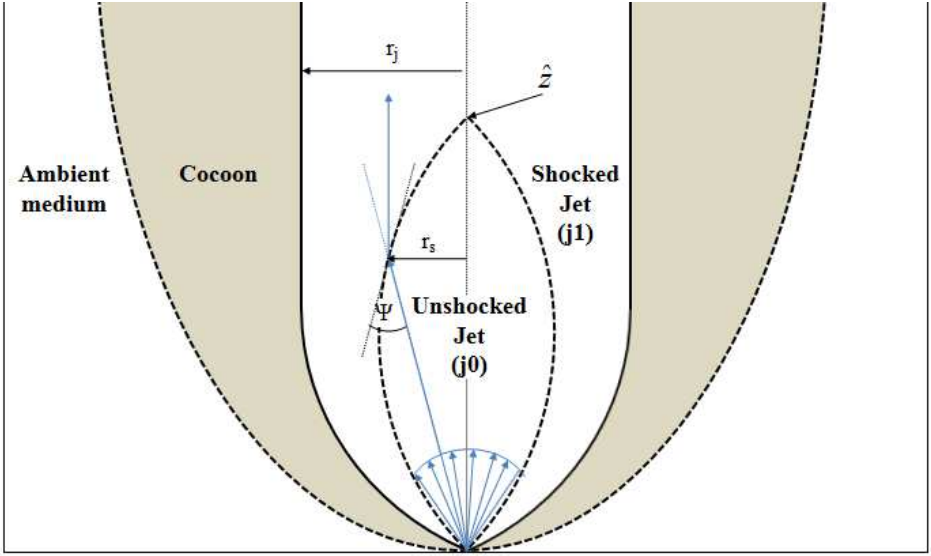}
  \caption{A schematic description of the collimation shock in the collimated
  regime. The collimation shock is marked in a dashed black line, and it separates the jet into
  an un-shocked region ($j0$) and a shocked region ($j1$).
  The jet flow-lines, marked with light blue arrows, are initially radial.
  They intersect the shock at point $(r_s,z)$ and form an angle $\Psi$ with the shock's surface.
  Downstream of the shock the flow-lines are deflected toward the jet's axis.
  The shock crossing point of a specific flow-line is illustrated.
  The shock converges onto the jet's axis at $\hat z$.
  Above $\hat z$ the jet maintains a constant cylindric radius, $r_j$.}\label{shock_fig}
\end{figure}

In the collimated regime the pressure in the cocoon is
sufficiently strong to compress the jet and form a strong oblique
shock at the base of the jet. The jet's head is then at full causal
contact and therefore $\eta=1$. The collimation shock divides the
jet into a pre-shocked and a shocked region. We designate them by
subindices $j0$ and $j1$ respectively  (See fig. \ref{shock_fig}).
The flow-lines in area $j0$ are radial and they encounter the shock
at an angle $\Psi(z)$, defined as the angle between the original
direction of the flow-line and the shock surface. As the flow-lines
pass through the shock, they are deflected toward the jet's axis.
Some of the flow energy is dissipated generating the pressure that
supports the jet against the compression of the cocoon. The shock's
geometry is set by the balance between the cocoon's pressure and the
upstream momentum flux, normal to the shock, that enters at the
upstream \citep[][]{KomisFall97,BL07}:
\begin{equation}\label{coll_shoc_jump}
\rho_{j0}c^2h_{j0}\Gamma_{j0}^2\beta_{j0}^2\sin^2\Psi+P_{j0}=P_c.
\end{equation}
Since the jet is relativistic and its ram pressure is much larger than
its internal pressure, $P_{j0}$ can be neglected on the LHS of
eq. (\ref{coll_shoc_jump}).
The radial geometry of the unshocked flow-lines implies that
$\rho_{j0}c^2h_{j0}\Gamma_{j0}^2\beta_{j0}^2\propto z^{-2}$.
Therefore to keep a uniform pressure at the downstream
$\Psi$ must increases with $z$. Consequently, the shock curves
toward the axis until it converges on it  at some point.
In the small angle approximation, to a first order,
$\sin\Psi=\left(\frac{r_s}{z}-\frac{dr_s}{dz}\right)$, where
$r_s$ is the cylindrical radius of the collimation shock (see fig. 2).
In this limit eq. (\ref{coll_shoc_jump}) becomes
a first order ODE whose solution is
\citep[][]{KomisFall97,BL09}:
\begin{equation}\label{rs}
r_s=\theta_0z\left(1+Az_*\right)-A\theta_0z^2,
\end{equation}
where $A\equiv\sqrt{\frac{\pi cP_c}{L_j\beta_{j0}}}$, and $z_*$ is
the hight where the jet is first affected by the cocoon and the
collimation shock forms. As long as the pressure in the jet is
larger than the pressure in the cocoon, the jet is unaware of the
cocoon. As the jet accelerates pressure is gradually converted to
kinetic energy until eventually it reaches a point where
$P_{j0}(z)=P_c$. At this point the jet's compression by the cocoon becomes
significant. Therefore if the jet-medium interaction begins at the
injection point $z_*$ can be tracked to the point where
$P_{j0}=P_c$. Alternatively, if the jet is injected into a cavity of
radius $R_*$, $z_*=\max\{R_*, z${\scriptsize $(P_{j0}=P_c)$}$\}$. Note
that once the shock forms, its geometry depends only on the jet
luminosity and it is the same for an accelerating or a
non-accelerating jet. The shock converges at $r_s=0$, namely at
\begin{equation}\label{z*}
\hat z=A^{-1}+z_*.
\end{equation}
As long as $z_*<A^{-1}$ it can be neglected
from eqn. (\ref{rs},\ref{z*}), which implies that $R_*$ must be smaller
than $A^{-1}$ as well.
When $z_*>R_*$ we can use the jet luminosity at $z_*$ which
holds $L_j\geq4\pi P_{j0}\Gamma_{j0}^2\beta_{j0}z_*^2\theta_0^2c$ together
with the condition that $P_{j0}=P_c$ and extract $z_*$.
Substituting $A$, we then get that $z_*<A^{-1}$ if
\begin{equation}\label{eq:fast}
\Gamma_{j0}(z_*)\beta_{j0}(z_*)\gtrsim\theta_0^{-1}.
\end{equation}
We define a ``sufficiently fast" jet as one that satisfies this
condition. In this situation we can ignore $z_*$ as long as $R_*<A^{-1}$. Since we work in
the limit where $\theta_0$ is small, eq. (11) also implies that the
jet is relativistic and
therefore in the following we approximate $\beta_{j0} = 1$ and
$\beta_{j1} = 1$. Note that as long as $\Gamma_{j0}(z)\leq\theta_0^{-1}$
the jet expands sideways under its own pressure and its
opening angle increases as it propagates. In such a case it is meaningless to
discuss a constant initial opening angle, and our model does not
hold. The jet stop expanding when the angle of the flowlines is
of the order of $\Gamma_{0}^{-1}$, one over the injected Lorentz factor.
Beyond this point it corresponds with our
condition for neglecting $z_*$. In this case we can approximate
$\theta_0$ to be $\sim\Gamma_{0}^{-1}$ and neglect $z_*$.

To estimate the jet's cross-section we approximate the jet to be
conical up to the point where collimation has a sizable effect on
the jet's geometry, which is where the collimation shock is
roughly parallel to the $z$ axis.
From eq. (\ref{rs}) this occurs at $\hat z/2$. 
Above this point the jet is collimated and since the cocoon
pressure is roughly uniform, the jet cross-section doesn't change
much, and it can be taken to be constant. In this approximation the
cylindric radius of the jet is $r_j(z>\hat z)\simeq \hat
z\theta_0/2$ and its cross-section is:
\begin{equation}\label{sigma}
\Sigma_j(z>\hat z)\simeq\frac{1}{4}\pi\hat z^2\theta_0^2\simeq\frac{L_j\theta_0^2}{4cP_c}.
\end{equation}
The luminosity above $\hat z$ is
\begin{equation}\label{Lj}
L_j\simeq4P_c\Gamma_{j1}^2\Sigma_jc.
\end{equation}
Substituting $L_j$ in eq. (\ref{sigma}) we obtain the
Lorentz factor of the shocked jet material above $\hat z$:
\begin{equation}\label{Gamma_j}
\Gamma_{j1}(z>\hat z)=\frac{1}{\theta_0}.
\end{equation}
This implies that, as long as the jet is sufficiently fast at the injection point, the Lorentz factor of the collimated jet
depends only on the injection angle, and it is independent of the initial Lorentz factor.

The remaining system parameters are calculated by substituting eq.
(\ref{sigma}) for the jet's cross section into equations (3-7).
Eqn. (\ref{tild_L}) and
(\ref{Pc_ann}) yields:
\begin{equation}\label{tilde_L_strong}
\tilde L=\frac{4}{\theta_0^2}\frac{P_c}{\rho_ac^2}\simeq
\frac{4}{c^2} t^{-4/5}L_j^{2/5}\rho_a^{-2/5}\theta_0^{-8/5},
\end{equation}
and
\begin{equation}\label{Pc_ann1}
P_c\simeq t^{-4/5}L_j^{2/5}\rho_a^{3/5}\theta_0^{2/5}.
\end{equation}
The head's position  is calculated using $z_h\sim\beta_ht$,
and the linearized expression for $\beta_h$ given in  eqn. (4a,b):
\begin{equation}\label{zh_app}
z_h\sim~\left\{
\begin{array}{ccc}
\tilde L^{1/2}ct& , & \tilde L\ll1\\
 & \\
ct & , & 1\ll\tilde L<\theta_0^{-4/3}.
\end{array}
\right.
\end{equation}
Taking the cocoon's cylindric radius
$r_c\sim t\sqrt{P_c/\bar\rho_a}$ and the jet's cylindric radius $r_j\sim\theta_0\sqrt{L_j/cP_c}$,
we calculate the opening angle of the cocoon and the jet respectively:
\begin{equation}\label{thta_c_app}
\theta_c\equiv\frac{r_c}{z_h}\sim~\left\{
\begin{array}{ccc}
\theta_0 & , & \tilde L\ll1\\
 & \\
\tilde L^{1/2}\theta_0 & , & 1\ll\tilde L<\theta_0^{-4/3},
\end{array}
\right.
\end{equation}
and
\begin{equation}\label{thta_j_app}
\theta_j\equiv\frac{r_j}{z_h}\sim~\left\{
\begin{array}{ccc}
\tilde L^{1/4}\theta_0^2 & , & \tilde L\ll1\\
 & \\
\tilde L^{3/4}\theta_0^{2} & , & 1\ll\tilde L<\theta_0^{-4/3}.
\end{array}
\right.
\end{equation}
Note that if the head is non-relativistic, the cocoon's aspect ratio is constant and it equals
the jet's injection angle, up to a constant of order unity (see Appendix B).

We can now infer the parameter regime for which the jet is
collimated by the cocoon.
The jet is considered collimated if the
collimation shock converges below the jet's head. Therefore, the
transition to the un-collimated regime occurs when $\hat z=z_h$. We
find that this equality can take place only when $\tilde L>1$, thus
we substitute $t=z_h/c$ in eq. (\ref{Pc_ann1}) and get that for
$z_h\geq\hat z$, $P_c\gtrsim\rho_ac^2\theta_0^{2/3}$. Substituting
$P_c$ in eq. (\ref{tilde_L_strong}),
we find,  up to a constant of order unity, that the condition for
collimation is:
\begin{equation}
\tilde L\lesssim\theta_0^{-4/3}.
\label{eq:collimation}
\end{equation}
The numerical factor missing in this equation
depends on the density profile (see appendix B). This condition
corresponds to a limit on the head's Lorentz factor
\begin{equation}
\Gamma_h\lesssim\theta_0^{-1/3} .
\label{eq:headspeed}
\end{equation}
If the head accelerates beyond $\theta_0^{-1/3}$
the collimation shock fails to converge below the head, and
the jet can be approximated as having a conical shape.

The jet can shift from a collimated to an un-collimated state and
vise versa, depending on the density profile of the ambient medium.
According to eqn. (\ref{tilde_L_strong},\ref{zh_app}), $\tilde
L\propto P_c/\rho_a\propto \left(z^{-4}\rho_a^{-2}\right)^{\delta}$,
where $\delta=1/3$ if $\tilde L\ll1$ and $\delta=1/5$ if $1\ll\tilde
L<\theta_0^{-4/3}$. If the density profile is
steeper than $z^{-2}$, then $\tilde L$ increases with $z$ and the jet's
head accelerates. In such a case $\theta_j=\theta_0\frac{\hat z}{2z_h}$
increases with time. When $\tilde L\simeq\theta_0^{-4/3}$, the
collimation shock converges at the head, and $\theta_j=\theta_0/2$.
At higher values of $\tilde L$ the shock opens up and the jet
becomes conical. In the limit of $\rho_a\propto z^{-2}$, $\tilde L$
is constant and the jet's head velocity is constant. Eq.
(\ref{thta_j_app}) shows that in this limit $\theta_j$ is constant
as well, implying that $r_j\propto z_h$.
This makes an interesting case where the jet expands sideways at the same
rate it propagates upward. Although the opening angle of the jet,
$\theta_j\ll\theta_0$, it remains constant like in the case of a
conical jet.

\subsection{The un-collimated regime}

When  $\tilde L\gg \theta_0^{-4/3}$ ($\Gamma_h\gg\theta_0^{-1/3}$)
the pressure in the cocoon is
too weak to significantly affect the geometry of the jet. In this
case the jet remains conical to a good approximation and
\begin{equation}\label{sigma_weak}
\Sigma_j(z_h) = \pi z_h^2\theta_0^2.
\end{equation}
The collimation shock remains at the edge of the jet, resulting in a
coaxial jet structure of a cold, fast, inner spine surrounded by a
hotter and denser layer of the shocked jet material moving with  a
lower Lorentz factor (see fig. 1). The layer of shocked material
becomes thinner and thinner at higher values of $\tilde L$.  In this
regime most of the jet's plasma streams freely all the way to the
head (region $j_0$ extends to the head), and dissipates all its
energy in the reverse shock. The jet is therefore cold below the
head with some brightening at its limb, as opposed to a collimated
jet which is hot below the head, since its plasma is first shocked
much closer to the base by the collimation shock.
Substituting $\Sigma_j$ in eq. (\ref{tild_L}) and taking $z_h=ct$ gives:
\begin{equation}\label{tilde_L_conic}
\tilde L=\frac{L_j}{\pi\theta_0^2\rho_at^2c^5}.
\end{equation}
This implies that also
in the un-collimated regime $\rho_a\sim z^{-2}$ remains the limiting
profile which distinguishes between an accelerating and a decelerating head.

The cocoon has some distinctive properties which can be
classified to three subcases, according to the values of $\tilde L$
and $\eta$, the fraction of energy that flows from the head into the
cocoon (see also table 1). In each subcase the conditions in the
head are different, and this affects the amount of energy that flows out of the
head:
\begin{itemize}
\item[]{1)} $\theta_0^{-4/3}\ll\tilde L\ll\theta_0^{-4}$ ($\theta_0^{-1/3} \ll \Gamma_h \ll \theta_0^{-1}$):
The head is sufficiently slow. Different regions of the head
are in a causal contact with each other and all the energy in the
head can flow into the cocoon, thus $\eta\simeq1$.
The cocoon's pressure is calculated by substituting $\tilde L$ from eq. (\ref{tilde_L_conic}) in eq. (\ref{Pc_ann})
and using $z_h=ct$:
\begin{equation}\label{Pc_weak}
P_c\simeq t^{-1/2}L_j^{1/4}\rho_a^{3/4}\theta_0^{1/2}c^{3/4}\simeq\tilde L^{1/4}\theta_0\rho_ac^2.
\end{equation}
Under this pressure the temperature at the outer cocoon remains sub-relativistic
and the different parts of the cocoon maintain causal connection in the lateral direction.
As $\tilde L$ increases $P_c/\rho_a c^2$ grows and it approaches unity as
$\tilde L\rightarrow\theta_0^{-4}$.
In this limit the pressure becomes mildly relativistic and it pushes the edge
of the cocoon to a velocity $\beta_c\rightarrow1$, which is
above the local sounds speed, $c/\sqrt{3}$.
This results in a loss of causality in the transverse direction
which implies that the approximation of a uniform pressure no
longer holds. In addition the cocoon's aspect ratio approaches unity and
it can no longer be considered as cylindric.
Therefore above the limit of $\tilde L\simeq\theta_0^{-4}$ our model can only provide
the total energy in the cocoon.
This has no effect on the conditions in the jet and the jet's head, which
are well described by our model, since in
the un-collimated regime the jet is not sensitive to the pressure in the cocoon.

\item[]{2)} $\theta_0^{-4}\ll\tilde L\ll\Gamma_{j0}^4$ ($\theta_0^{-1}<\Gamma_h<\Gamma_{j0}$):
At this limit the reverse shock is still strong and pressure in the
head is large, but the jet's head loses causal connectivity in the
transverse direction.
Energy can flow to the cocoon only
from a thin annulus on the edge of the head with an opening angle
$\Gamma_h^{-1}$, which corresponds to a fraction
$\eta=2(\Gamma_h\theta_0 )^{-1}$ of the total energy that enters the
head.
Most of the energy that flows into the head remains trapped and accumulates
at a rate of $L_j\Gamma_h^{-2}$.
The total energy that flows into the cocoon can thus be estimated as
\begin{equation}
E_c=\eta L_j\Gamma_h^{-2}t\simeq\tilde L^{1/4}\theta_0\rho_ac^5t^3,
\end{equation}
where we substitute $L_j$ from eq. (\ref{tilde_L_conic}) and used the relation $\tilde L\simeq4\Gamma_h^4$
from eq. (4b).
Since $\beta_c=1$, the cocoon has a spherical rather than cylindrical
shape and it occupies a volume $V_c\simeq (ct)^3$.

\item[]{3)} $\Gamma_{j0}\ll\tilde L$ ($\Gamma_h\simeq\Gamma_{j0}$):
In this regime the reverse shock becomes Newtonian while the forward shock
remains relativistic and moves with a Lorentz factor $\simeq\Gamma_{j0}$.
This leads to different sound speeds in the head, above and below the contact discontinuity.
Below, within the shocked jet material, the sound speed $\ll c$. Above, within the shocked medium,
the sound speed is $c/\sqrt{3}$.
As a result the energy that flows into the cocoon comes mostly from the ambient medium part of the
head and it flows into the outer cocoon, while the inner cocoon which is fed by the shocked jet
material in the head becomes insignificant.
To measure the energy that enters the cocoon we first estimate the
energy per unit time that flows into the head through the forward shock:
$\dot{E}_{ha}=\rho_a\Gamma_{j0}^2\theta_0^2t^2c^5$.
The total energy in the cocoon is therefore:
\begin{equation}
E_c=\eta\dot{E}_{ha}t\simeq\Gamma_{j0}\theta_0\rho_at^3c^5,
\end{equation}
where $\eta=2(\Gamma_{j0}\theta_0)^{-1}$ in this case.
\end{itemize}

Table. \ref{tb:summ} below summarizes the different characteristics of the jet-cocoon in the four different collimation  regimes.

\subsection{Comparison with previous analytical works}

The propagation and interaction of a jet with an ambient medium
was studied in the past in various parameter regimes.
Here we briefly comment on some relevant works and discuss their
compatibility to our results.

The main new feature introduced in this work is a proper closure of the
set of equations using the conditions at the collimation shock.
This allows us to determine if the jet is collimated or not, given
its luminosity, initial opening angle and the ambient density.
When the jet is collimated it also allows us to determine its cross section,
$\Sigma_j$.
Earlier works differ from ours in the way they deal with this issue.
Most ignore this collimation shock and assume a specific value or functional
form for one of the variables (e.g. the jet cross section), obtaining a closure in this way.
Such solutions typically agree with ours if we substitute the value we
obtain from the full system of equations instead of the free variable
in these solutions.

\citet{BegelCiof89} analyzed the propagation of a non-relativistic galactic
jet in the inter galactic medium, assuming that the jet is collimated
and that it has a constant velocity $v_j<c$.
When using our expression for $\Sigma_j$ (eq. \ref{sigma}) in their
solution
(and taking the limit where the $v_j\rightarrow c$), their results agree
with ours for $\tilde L\ll1$.
\citet{Matzner03} extended this model to
relativistic jets.
Like \citet{BegelCiof89} he did not model the collimation
of the jet and used $\theta_j$ (which is related to $\Sigma_j$) as a given parameter.
This solution is consistent with ours (for $\tilde L\ll1$)
upon substitution of $\theta_j$ from our solution.
However, \citet{Matzner03} assumed that the jet is in fact conical,
with $\theta_j=\theta_0$, which is inconsistent in this limit of $\tilde L\ll1$.

\citet{LazzBeg05}, considered the collimation of the jet
by the cocoon's pressure. However they ignored the dissipation in the
collimation shock. Instead they assumed that the jet material expands
adiabatically, leading to a relation $\Gamma_{j}\propto\Sigma_j^{1/2}$.
As we show in sec. 3.1 (eqn. \ref{sigma}-\ref{Gamma_j}),
the collimation shock renders this relation invalid.
Consequently, the solution they obtained differs from ours
in all regimes.
Their solution has a smaller opening angle and a larger value of
$\Gamma_j$ at breakout.
Later on, \citet{Morsony07} attempted to calculate the geometry of the
collimation shocks.
But they have used an incorrect
expression for the momentum flux that crosses this shock, which led to a shock
that never converge to the axis ($\hat z\rightarrow\infty$).

Finally, \citet{MW01} analyzed the propagation of an
uncollimated GRB jet in the outer envelope of a red supergiant.
They considered only the properties of the jet's head, ignoring the surrounding cocoon.
Their solution for $\Gamma_h$ is valid for
$\theta_0^{-4/3}\ll\tilde L\ll\Gamma_{j0}^4$.

\begin{table}[!h]
\caption{The system's characteristics in the collimated and un-collimated regimes}
\vspace{2mm}
\begin{tabular}{|c|c|c||c|c|c|}
\hline
 & \multicolumn{2}{|c||}{\multirow{2}{*}{Collimated jet}} &
 \multicolumn{3}{|c|}{Un-collimated jet} \\
 \cline{4-6}
& \multicolumn{2}{|c||} {} & Causal head & Uncausal head & Free expansion \\
\hline
  & $\tilde L<1$ & $1\ll\tilde L<\theta_0^{-4/3}$ &
  $\theta_0^{-4/3}<\tilde L\ll4\theta_0^{-4}$ & $4\theta_0^{-4}<\tilde L\ll4\Gamma_{j0}^{4}$ &
  $\Gamma_j^{4}\ll\tilde L$ \\
 \hline\hline
$\tilde L$ & \multicolumn{2}{|c||}{$\left(\frac{L_j}{\rho_at^2\theta_0^4c^5}\right)^{2/5}$} &
 \multicolumn{3}{|c|}{$\frac{L_j}{\rho_at^2\theta_0^2c^5}$} \\
\hline
 $\beta_h$ & $\tilde L^{1/2}$ & 1 & \multicolumn{3}{|c|}{1} \\
 \hline
 $\Gamma_h^2$ & 1 & $\frac{1}{2}\tilde L^{1/2}$ &
 \multicolumn{2}{|c|}{$\frac{1}{2}\tilde L^{1/2}$} &
 $\Gamma_j^2$ \\
 \hline
 $\theta_j$ & $\tilde L^{1/4}\theta_0^2$ & $\tilde L^{3/4}\theta_0^2$ &
  \multicolumn{3}{|c|}{$\theta_0$} \\
  \hline
 $\Gamma_{j}^{\dag}$ & \multicolumn{2}{|c||}{$\Gamma_{j1}=\theta_0^{-1}$} &
\multicolumn{3}{|c|}{$\Gamma_{j0}$} \\
  \hline
 $P_c$ & \multicolumn{2}{|c||}{$\tilde L\theta_0^2\rho_ac^2$} &
\multicolumn{2}{|c|}{$\tilde L^{1/4}\theta_0\rho_ac^{2~~\dag\dag}$}
&
$\Gamma_j\theta_0\rho_ac^{2~~\dag\dag}$ \\
  \hline
 $\beta_c$ & \multicolumn{2}{|c||}{$\tilde L^{1/2}\theta_0$} &
 $\tilde L^{1/8}\theta_0^{1/2}$ &
\multicolumn{2}{|c|}{1} \\
  \hline
 $\eta$ & \multicolumn{2}{|c||}{1} & 1 &
 $\left(\tilde L^{1/4}\theta_0\right)^{-1}$ &
  $\left(\Gamma_{j0}\theta_0\right)^{-1}$\\
 \hline

  \multicolumn{6}{l} {\footnotesize All quantities are missing order
of unity constants of integration over the density profile. In case of }\\
   \multicolumn{6}{l} {\footnotesize   a power-law density profiles these constants can be
  calculated analytically and they are given in the }\\
 \multicolumn{6}{l} {\footnotesize  appendix.}\\
 \multicolumn{6}{l} {\footnotesize  $\dag$ $\Gamma_{j}$ is the jet Lorentz factor just below the head.}\\
 \multicolumn{6}{l}{\footnotesize $^{\dag\dag}$ This quantity represents the total energy in the cocoon
 divided by the volume $V_c=(ct)^3$}\\
 \multicolumn{6}{l}{\footnotesize   and not the real pressure, which is not calculated in this regime.}

\end{tabular}
\label{tb:summ}
\end{table}

\section{Comparison with numerical simulations}

\begin{figure}[!h]
  \includegraphics[width=5in]{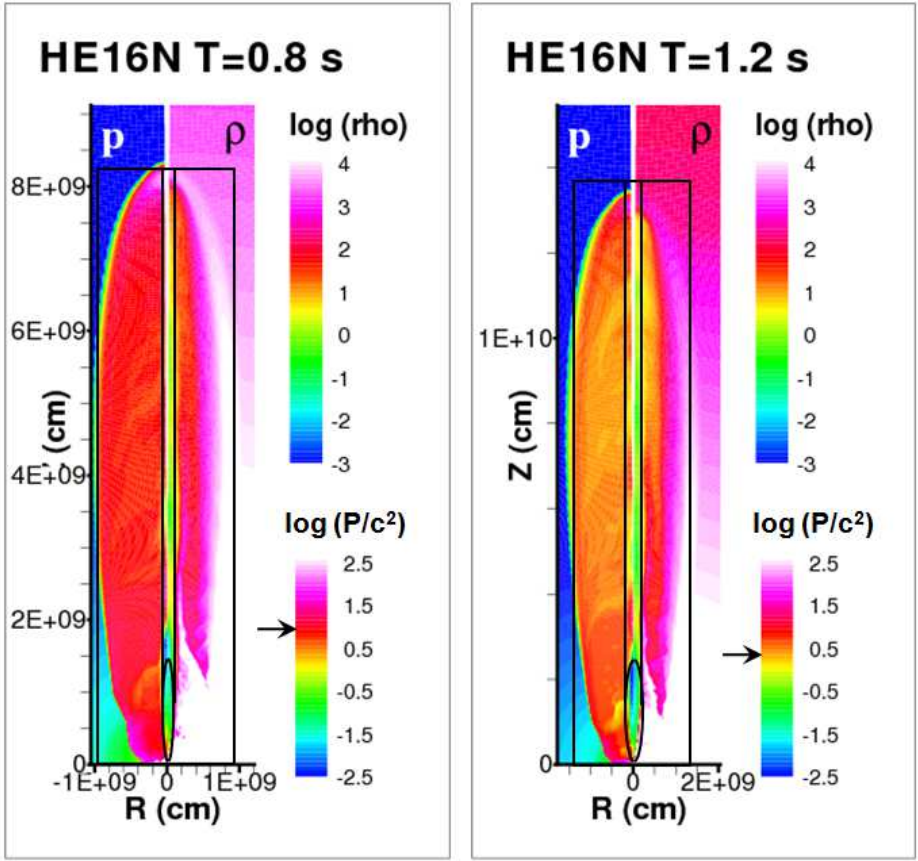}
  \caption{The calculated size of the jet, the cocoon and the collimation shock
  drawn in black lines on top of a simulated jet by \citet{Mizuta09}.
  The left (right) panel shows a snapshot of the jet and the cocoon after 0.8(1.2) sec.
  Color codes are of equal pressure divided by $c^2$ on the left, and equal density on the right.
  The black arrow shows the average value of the cocoon's pressure from our calculation.
  Figures are at curtesy of A. Mizuta.}\label{shock_fig1}
\end{figure}

Jet simulations have been carried out extensively by various authors.
We turn now to compare our results with two recent numerical simulations.
First, we consider the simulation by \citet{Mizuta09} who modeled the
propagation of a relativistic jet in the envelope of a massive star.
To model the star they used a numerically calculated density profile
from \citet{Woosley06}  \citep[model number HE16N in][]{Mizuta09}.
This star has a radius of $\sim6\times10^{10}$
cm, and it has a density profile which can be divided into three
parts. The inner part (up to $\sim1.2\times10^{10}$ cm) has an
average powerlaw profile with an index $\alpha=-2.5$. Above it
(up to $4\times10^{10}$ cm) the profile is steeper with an averaged index
$\alpha\simeq-4.5$, and it drops sharply from there to the edge of the
star. The jet in the simulation is hot and it is injected with
a Lorentz factor, $\Gamma_{0}=5$ into a cone of an opening angle $\theta_0=5^\circ$
and an initial altitude of $z_*=10^8$ cm.
Since $\theta_0<\Gamma_{0}^{-1}$ the jet goes
through an initial transitory phase where it's opening angle
increases until it stabilizes when
$\theta_{j0}(z)\simeq\Gamma_{j0}^{-1}(z)\simeq10^\circ$.
We therefore take in our calculations $\theta_0=10^\circ$.
Figure (\ref{shock_fig1}) shows
two snapshots of the jet and the cocoon at times 0.8 sec and 1.2 sec
after the injection\footnote {$^{,3}$ The snapshots are a curtesy of
A. Mizuta, and are taken from a simulation published in
\citet{Mizuta09}.}. The times were chosen so that the jet is
far enough from the injection point and the opening angle at $z_*$
is already stable. The color codes represent equal values of the
pressure normalized by $1/c^2$ (left side) and density (right side).
Our calculation of the jet, the cocoon and the collimation shock is
drawn in black lines on each panel. Due to the relatively large $z_*$
(see below), 
we did not use the analytic formulae, instead  we have integrated
eqn. (2,5,6,10,12)  numerically to obtain a solution.
It can be seen
that in a density profile of $\alpha=-2.5$, the approximation of a
uniform pressure in the cocoon fits well with the pressure profile
in the simulation. Our results agree with the numerical
results to a good accuracy.

\begin{figure}[!h]
  \includegraphics[width=5in]{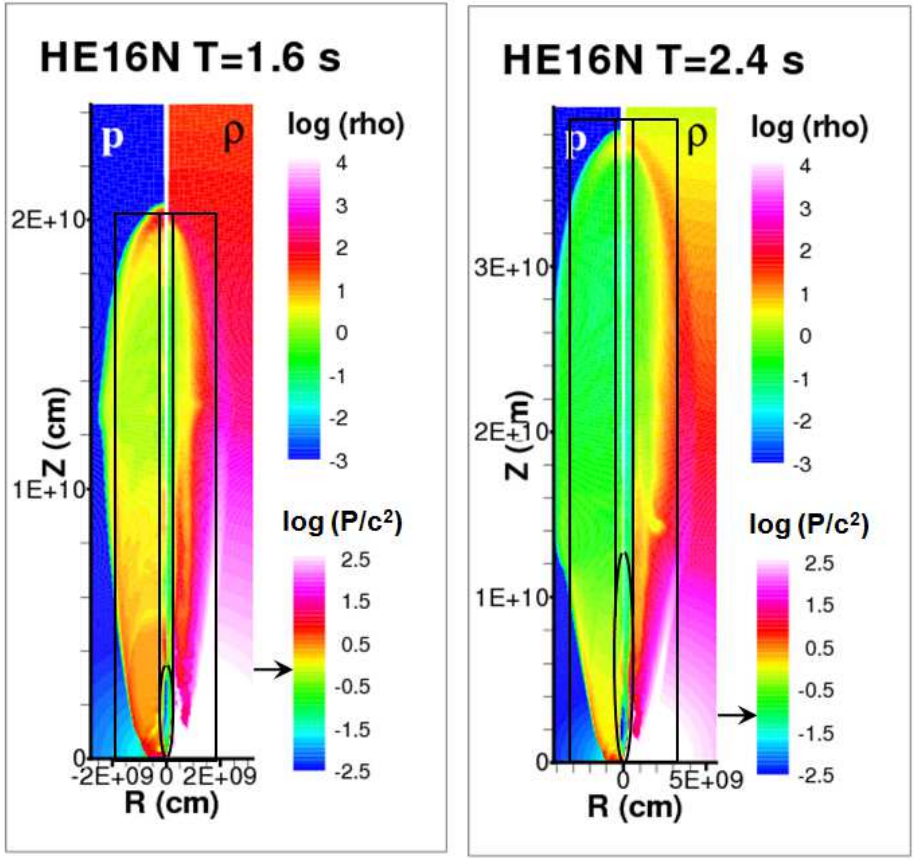}
  \caption{The calculated size of the jet, the cocoon and the collimation shock
  drawn in black lines on top of a simulated jet by \citet{Mizuta09}.
  The left (right) panel shows a snapshot of the jet and the cocoon after 1.6(2.4) sec.
  See fig. \ref{shock_fig1} for further details.
  Figures are at curtesy of A. Mizuta.}\label{shock_fig2}
\end{figure}

Figure (\ref{shock_fig2}) presents two snapshots of the jet and the
cocoon in the region with the steeper density profile, just before
the sharp drop$^3$. The steeper gradient in the stellar density
profile leads to an acceleration of the jet's head, which changes
the rate of energy flow into the cocoon. Nevertheless it can be seen
that the pressure remains uniform to a good approximation through
most of the cocoon's hight, due to the survival of the contact
discontinuity that keeps the inner cocoon intact.
The pressure calculated by our model fits well with the simulated pressure in the upper, uniform, part of the cocoon.
At $\sim1.2\times10^{10}$ cm the contact discontinuity becomes unstable and below this point
the light matter from the inner cocoon mixes with the
heavy material in the outer cocoon.
As a result the matter in the bottom part of the cocoon is no longer in contact with
the matter in the upper pars and a pressure gradient is formed.
As a consequence the collimation shock converges at a lower
altitude than what we calculate (at $\sim4\times10^9$ cm), which leads to a jet that is narrower at its base.
But due to the negative pressure gradient the jet widens to a similar size as calculated
by our model, and the jet's head has similar properties (width and velocity) as in our calculation.
The region where the contact is destroyed corresponds to a sharp step in the stellar density profile
that is located at $\sim1.2\times10^{10}$ and separates the shallow profile of $\alpha=-2.5$
from the steeper one with $\alpha=-4.5$. At this step the growth rate of instabilities on the contact
discontinuity increases.

\newpage

We also compared our model with the simulations of
\citet{Zhang03}. They examined the consequences of changing the
injection angle and the luminosity of the jet on the propagation in
a stellar mantle. Table \ref{table_zhang} shows the initial
parameters and the breakout times of the jet from the stellar
surface in the three cases examined. In the right column we added
the breakout times calculated by our analytical model, using the
same stellar density profile and initial jet parameters . It
can be seen that our results agree to within $10\%$ with the simulated
ones.

When comparing our results with numerical simulations we should
recall that the these simulations usually use a relatively
large value of $z_*$, the radius in which the jet is injected to the stellar envelope.
Such a jet is initially wider than a similar
jet that is injected at a smaller radius and at this stage
it propagates slower. Eventually the solution converges but the
resulting breakout time is longer by 
$\simeq {z_*}/{c}\sqrt{\frac{\pi z_*^2\theta_0^2\rho_a c^3}{L_j}}$, which
becomes significant for $z_*\gtrsim10^{-3}R$, where $R$ is the stellar radius.
The values of $z_*$ in jet simulations are usually larger than that and therefore
we integrate numerically
eqn. (2,5,6,10,12) to calculate the breakout times of the jet presented in table 2.

\begin{table}[h]
\caption{Breakout Times}\label{table_zhang}
\vspace{2mm}
  \centering
  \begin{tabular}  {c c c c c}
  \hline\hline
  Model$^a$ & $L_j/10^{51} erg$ & $\theta_0$ & $t^b$ [sec] & $t^c$ [sec] \\
  \hline
  JA & $1.0$ & $20^\circ$ & 6.9 & 6.3 \\
  JB & 1.0 & $5^\circ$ & 3.5 & 3.8 \\
  JC & 0.3 & $10^\circ$ & 5.7 & 5.2 \\
  \hline
   \multicolumn{5}{l}{\footnotesize $^a$ Data from \citet{Zhang03}}\\
   \multicolumn{5}{l}{\footnotesize $^b$ The breakout time according to the simulation}\\
   \multicolumn{5}{l}{\footnotesize $^c$ The breakout time according to the analytic calculation.}

 \end{tabular}
\end{table}

Though our analytical model greatly simplifies the conditions in the jet
and the cocoon, it shows a remarkable agreement with the results of the simulations presented above, which
is slightly better than what we would expect in the general case.
Generally we expect an order of unity agreement for all density
profiles that are not too steep, so that the forward and reverse
shock of the head remain in causal contact and that the
approximation of uniform cocoon pressure is reasonable. As we see in
the comparison to \citet{Mizuta09} even a power-law density gradient
as steep as $\rho_a \propto z^{-4.5}$ satisfies these conditions.
These conditions break down however in a very steep density profile, such
as the one at the edge of a stellar envelope, where the forward
shock accelerates and lose causal contact with the jet.

\section{Jet collimation in astrophysical environments}

According to our model when $\tilde L>\theta_0^{-4/3}$ the jet
is uncollimated, and it has a conical shape with $\theta_j=\theta_0$.
It implies that in this regime
$\tilde L=\frac{L_j}{\pi z_h^2\theta_0^2\rho_ac^3}$.
When $\tilde L<\theta_0^{-4/3}$, on the other hand, the jet is collimated and
$\theta_j<\theta_0$.
We can therefore formulate the condition for collimation as:
\begin{equation}\label{eq:collim_cond}
\frac{L_j}{\pi z_h^2\rho_a c^3}<\theta_0^{2/3},
\end{equation}
and evaluate the conditions in the jet that lead to its collimation
in different astrophysical media.

The conventional view about long GRBs is that they originate from
massive stars that collapse \citep[e.g.][]{Woosley93}.
The jet propagation in this model is characterized by two stages:
pre-breakout and post-breakout.
Prior to the breakout the jet propagates inside the star which is
considered to be massive ($M\sim10M_{\odot}$) and rather
compact ($R\sim R_\odot$) Wolf Rayet star \citep[e.g.][]{Woosley06,Crowther07}.
The density profile in the stellar envelope can be approximated as
$\rho_a=\bar\rho(z/R)^{-\alpha}$, with $2\lesssim\alpha\lesssim3$
\citep[e.g.][]{MatzMck99}, and $\bar\rho=\frac{3-\alpha}{4\pi} MR^{-3}$ is the average
density.
Substituting this in eq. (\ref{eq:collim_cond}) we get that the jet is
collimated if:
\begin{equation}\label{eq:collim_star}
L_j\lesssim10^{54}\left(\frac{z_h}{R}\right)^{2-\alpha}\left(\frac{R}{R_\odot}\right)^{-1}
\left(\frac{\theta_0}{10^\circ}\right)^{2/3}\left(\frac{M}{10M_{\odot}}\right)~erg/s.
\end{equation}
The luminosity we measure in GRBs reflects the luminosity of the jet after it
breaks out. We take here the simple approach that this luminosity is constant
over time.
Under this assumption we can
use the observed jet luminosities to estimate the collimation of the jet inside the star.
Correcting to a typical beaming angle of
$5^\circ-10^\circ$, the observed values give $L_j\lesssim10^{52}$ ergs/s.
In addition, as we show below, the measured values of the beaming
angle of the GRB are similar to the size of the injection angle at the base of the jet, $\theta_0$.
Substituting this in eq. (\ref{eq:collim_star}) we get that GRB jets are collimated before they breakout.
The propagation of the jet inside a stellar envelope, and the implications on
GRB observations are discussed to a greater
extent in \citet{Bromberg11}.

Once the jet breaks out its opening angle, $\theta_j$, can be measured,
for example by identifying a "jet break" in the afterglow lightcurve.
Therefore it is better to express the
condition for collimation (eq. \ref{eq:collim_cond}) in terms of $\theta_j$ rather than $\theta_0$.
Generally $\theta_j\leqslant\theta_0$ and $\theta_j = \theta_0$ if the
jet is uncollimated.
Therefore if
$\tilde{L}>\theta_j^{-4/3}$ it follows that $\tilde
L>\theta_0^{-4/3}$. On the other hand if $\tilde{L}<\theta_j^{-4/3}$ then
necessarily $\tilde L<\theta_0^{-4/3}$ (otherwise the jet is uncollimated
and $\theta_j=\theta_0$). Thus eq. (\ref{eq:collim_cond}) can
be used with $\theta_j$ replacing $\theta_0$.

The medium outside the star can either be a dense wind ejected
from the surface of the star, having a typical
density profile $\rho_a=a^*/z^2$, where $a^*\simeq5\cdot10^{11}$ g/cm,
or it can be the constant density ISM.
If the jet breaks out into a stellar wind environment 
it is collimated if:
\begin{equation}\label{collim_jet_lumin0}
L_j\leq10^{43}\left(\frac{\theta_j}{10^\circ}\right)^{2/3}\left(\frac{a^*}{5\cdot10^{11}~g/cm}\right)~erg/s.
\end{equation}
If alternatively it propagates in the ISM
it is collimated if
\begin{equation}\label{collim_jet_lumin1}
L_j\leq3\times10^{33}\left(\frac{z}{10^{13}~cm}\right)^2\left(\frac{\theta_j}{10^\circ}\right)^{2/3}\left(\frac{\rho}{10^{-24}
g/cm^{3}}\right)~erg/s.
\end{equation}
Since the observed luminosity is orders of magnitude higher than both of these
limits, the jet becomes un-collimated
when it propagates in these media.
Therefore, as the jet, which was collimated inside the star, breaks out,
it rapidly expands sideways and accelerates.
Without any external interference
the opening angle stabilizes when $\theta_j\simeq\Gamma_{j1}^{-1}$, where $\Gamma_{j1}$ is
the Lorentz factor of the jet at breakout, and since
$\Gamma_{j1}\simeq\theta_0^{-1}$ (eq. \ref{Gamma_j}) we get that
the opening angle after breakout should be $\simeq\theta_0$.
However the breakout of the cocoon, which occurs simultaneously
with the jet, limits the sideways expansion.
Thus the final opening
angle of the jet at early times may be smaller than $\theta_0$.
Note that the smaller opening angle only tightens the constraints
for collimation given in eqn. (\ref{collim_jet_lumin0}) and (\ref{collim_jet_lumin1}).
At late times, when the cocoon clears out, the jet is no longer confined
and maintains its initial injection angle.
Therefore measuring $\theta_j$ at late times
gives information about the conditions
at the injection site of the jet, even before the breakout when the
jet was still buried in the star and was in fact collimated.

Microquasars (MC) and X-ray binaries jets, have
luminosities of $\sim10^{39}$ erg/s, and their opening angles $\lesssim10^\circ$.
Using eq. (\ref{collim_jet_lumin1}) we get that these jets
may begin uncollimated, but beyond
$\sim2\cdot10^{-3}$ pc they become collimated due to the interaction with the ISM.
The collimation of MC and X-ray binaries jets was also proposed by \citet{Miller06},
who showed that if the values of $\Gamma_j$ are low ($\sim10$) it implies that
the jets are collimated.
Our calculation therefore supports this claim regardless of the Lorentz factor,
which can not be measured reliably at the present time.

Powerful Quasar jets extend to distances of hundreds of kpc from their sources,
and they propagate in a much thinner medium.
Their typical power is $<10^{47}$ erg/s, and their opening angle
$<10^\circ$. The density profile in the galactic hallo at such
distances is usually considered to be isothermal with $\alpha<2$,
and a mass density of the order of $\sim10^{-27}-10^{-28}$ g/cm$^3$
at a distance of $\sim10$ kpc
\citep[e.g.][]{Bulbul10,Capelo10}. Substituting these values in eq.
(\ref{collim_jet_lumin1}) we get that beyond this distance quasars jets are
hydrodynamically collimated by their cocoons.

\section{Summary}

In this work we present an analytical study of the propagation of a relativistic
hydrodynamic jet in an external medium with a general density profile.
The interaction of the jet with the medium results in the formation of a
shocked "head" at the front of the jet, and an over pressured cocoon with a
rather uniform distribution of energy surrounding the jet.
When the pressure of the cocoon, $P_c$, is larger than the internal pressure of the jet,
it compresses the jet and leads to the formation
of an oblique shock at the base of the jet.
Under some conditions, determined in this work, the shock converges to the
jet's axis, and the jet becomes collimated.
This changes the jet's properties, and affects the conditions in the cocoon.
Our model follows the evolution of the jet, the jet's head, the cocoon and the collimation shock,
given three initial parameters: the jet's luminosity, $L_j$, the
jet's opening angle at the injection point, $\theta_0$, and the density of the ambient
medium, $\rho_a(z)$.
We determine the condition for collimation, and provide a self-consistent,
time dependent, solution to the system's parameters that depends
on these three initial parameters alone.
The main results of our analysis can be
summarized as:
\begin{itemize}
\item The jet's evolution can be classified into two regimes: a collimated and an un-collimated,
according to the strength of interaction of the jet with the ambient
medium. The two regimes are distinguished by the parameter $\tilde
L$ and by $\theta_0$ (or $\theta_j$), which represents the ratio of the jet's energy density  to
the rest-mass energy density of the ambient medium at the
location of the head.
\item When $\tilde L<\theta_0^{-4/3}$ the interaction is strong and the jet is collimated.
In this regime
the collimation shock converges on the jet's axis at some point,
$\hat z$,  below the head.
Above $\hat z$ the jet is cylindrical to a good approximation and its
width is estimated as $\hat z\theta_0/2$.
This width implies that the  Lorentz factor of the collimated ejecta satiafies
$\Gamma_{j1}=\theta_0^{-1}$.
\item If $\tilde L\ll1$, the head of the collimated jet is non-relativistic ($\beta_h\ll1$).
In this limit the cocoon's expansion velocity
is proportional to the head's velocity, and it has a constant opening angle $\theta_c\simeq\theta_0$.
\item When $\tilde L\gg\theta_0^{-4/3}$
the collimation shock fails to converge and it remains at the edge of the jet.
The jet in this regime in un-collimated and it remains conical to a good approximation.
It has a coaxial structure
of a fast unshocked spine surrounded by a denser layer of shocked material with lower Lorentz factor.
\item When $\tilde L\gg\theta_0^{-4}$, $\Gamma_h>\theta_0^{-1}$, and the
head is not causal in the transverse direction. Here only a fraction
$\sim (\Gamma_h\theta_0)^{-1}$ of the jet energy flows into the cocoon.
This energy is enough to produce a mildly relativistic pressure in the cocoon, and
accelerate cocoon's edge to a velocity $\beta_c\sim1$.
\item When $\tilde L>4\Gamma_j^{-4}$ the reverse shock becomes Newtonian and the
head no longer affects the jet's propagation. Matter continue to
stream into the cocoon only from the forward shock, and it generates a
relativistic pressure in the cocoon.
\item The value of $\tilde L$ can change with time, even for a constant $L_j$ and $\theta_0$,
according to the slope of the density profile. In a density profile
$\rho_a\sim z^{-\alpha}$ with $\alpha>2$, $\tilde L$ increases with
time. This corresponds to an acceleration of the jet's head.
In such cases a collimated jet opens up and become un-collimated
once $\tilde L$ becomes larger than $\theta_0^{-4/3}$. The opposite
evolution take place when $\alpha<2$. In the case of $\alpha=2$,
$\tilde L$ is constant, the head maintains its velocity, and the
opening angle of a collimated jet remains constant.
\item The choice of the initial injection radius of the jet, $z_*$, 
can affect the jet's breakout time, where larger
values of $z_*$ increase this time.
This effect becomes important when the ratio of $z_*$ to the stellar radius 
$\gtrsim10^{-3}$.
Numerical simulations typically use values of $z_*$ which are above this limit
and therefore their resultant breakout times are affected by this choice.
\end{itemize}

Our model provides a general frame to study the properties of
relativistic jets in different media.
It can be used to examine various phenomena, such as the minimal energy
required by the jet to break out of a boundary surface at a finite
distance, like the edge of a star.
It can also be used to test the energy feedback into
the stellar envelope in the case of a GRB jet that penetrates a star,
or the IGM in the case of AGN jets. This may help understands
better issues such as the problem of IGM heating.
Our model confirms that GRB jets, which form inside a star, are collimated
before they break out and become uncollimated afterwards.
Microquasars and X-ray binary jets, however, may start uncollimated, but the
interaction with the ISM leads to their collimation beyond
$\sim2\cdot10^{-3}$ pc.
We also show that quasar jets are collimated as well at large
distances from their sources ($\gtrsim10$ kpc).
We stress that in our calculations we assume that the magnetic fields in the jet,
in the region where the jet undergoes the collimation and above, are dynamically
unimportant and therefore can be ignored.
High magnetization in this region will alter our results, for example by
preventing the formation of the collimation shock, but will keep the
other basic properties of the model, i.e. the formation of the cocoon and the
collimation of the jet.

We are in dept to A. Mizuta for providing us results from his numerical simulations.
We also thank Y. Lubarsky for useful discussions, and to the anonymous referee
for helpful comments.
O.\,B.\ and T.\ P. were supported by the Israel Center for Excellence for High Energy Astrophysics and
by an ERC advanced research grant. E.\,N.\ was supported in part by the Israel Science Foundation
(grant No.\ 174/08) and by an EU International Reintegration Grant.
R.\, S. \ is partially supported by IRG \& ERC grant, and a Packard fellowship.

\renewcommand{\theequation}{A-\arabic{equation}}
\setcounter{equation}{0}  
\section*{APPENDIX A - Testing the assumption of energy conservation in the jet}\label{sec_consistency}

In calculating the parameter $\tilde L$ and the velocity of the
head, we assume that the jet material does not lose energy to work
as it flows from the injection point to the head. Such an assumption
is natural if the jet is conical, since the jet has a constant
opening angle and its envelope at a given point doesn't expand with
time. However when the jet is collimated its width changes with
time, and the jet loses or gains energy due to mechanical work
preformed against the pressure of the cocoon. When $\tilde L\ll1$
the head moves at a sub-relativistic velocity. In this limit, to a
first order in $\beta_h$, the work done due to sideways expansion of
the jet is $dW=P_cdV=P_cz_hd\Sigma_j$. This work can be neglected as
long as it is much smaller than the energy added to the system:
$L_jdt$. We therefore define
$\epsilon_j=\frac{P_cz_h}{L_j}\frac{d\Sigma_j}{dt}$ as the relative
amount of energy lost to expansion of the jet. Our approximation
holds as long as $\epsilon_j\ll1$. Substituting  $\Sigma_j$ from eq.
(\ref{sigma}) and $P_c$ from eq. (\ref{Pc_ann1}) gives:
\begin{equation}
\epsilon_j\simeq\theta_0^2\beta_h=0.03\left(\frac{\theta_0}{10^\circ}\right)^2\beta_h.
\end{equation}
Implying that the assumption of negligible energy losses in the jet
is always valid in the limit of small injection angle.

When $1\ll\tilde L\ll\theta_0^{-4/3}$ the head is relativistic, and the energy flux
that enters the head through the reverse shock is reduced by a factor of $\sim\Gamma_h^{-2}$.
Since we are interested in the energy that enters the head we compare
the work lost to expansion with $L_j\Gamma_h^{-2}dt$.
This gives $\epsilon_j=\frac{\Gamma_h^2P_cz_h}{L_j}\frac{d\Sigma_j}{dt}$,
and applying  the same consideration as before we get that
\begin{equation}
\epsilon_j\simeq\Gamma_h^2\theta_0^2.
\end{equation}
In this regime $\Gamma_h<\theta_0^{-1/3}$ (see table 1), therefore
the relative energy lost to work in a relativistic collimated jet
maintains $\epsilon_j<\theta_0^{4/3}$. But in this regime of a
collimated jet with a relativistic head, the injection angle is
limited by $\theta_0^{4/3}\ll1$. This guarantees that
$\epsilon_j\ll1$, and implies that our approximation is always valid
in this type of jets as well.

\renewcommand{\theequation}{B-\arabic{equation}}
\setcounter{equation}{0}  
\section *{APPENDIX B - Analytic solutions to the relativistic and non-relativistic limits}

The system's behavior is determined by the equations (\ref{bh},
\ref{Pc}, \ref{bc}, \ref{sigma}). Generally these equations should
be solved numerically, due to the non-linearity of $\beta_h$. But in
the limits of $\tilde L\ll1$ and $\tilde L\gg1$, eq. (\ref{bh}) can
be linearized and the integration over time can be solved
analytically. We define the following integration parameters:
$\bar\rho_a(z_h)=\int\rho_adV/V\equiv\varrho\rho_a(z_h)$,
$z_h=\int\beta_hdt\equiv\zeta\beta_ht$,
$r_c=\int\beta_cdt\equiv\xi\beta_ct$ and and
$\int\Gamma_h^{-2}dt\equiv\varepsilon\Gamma_h^{-2}t$. Using these
parameters, eq. (\ref{Pc_ann}) can be written as:
\begin{equation}
P_c=\left(\frac{L_j\rho_a}{3\pi c}\right)^{1/2}\tilde L^{1/4}t^{-1}\left(\frac{\xi\varrho}{\zeta\varepsilon^2}\right)^{1/2}.
\end{equation}
Each of these parameters takes a different value when $\tilde L\ll1$ and when $\tilde L\gg1$.
If the density profile is a powerlaw of the sort
$\rho_a\sim z^{-\alpha}$, the parameters become constants and their value
depends on $\alpha$.
The resulting solutions including the integration parameters in each regime are presented below,
for convenience in presenting $\varrho$ we took density profiles with $\alpha<3$.

\subsection *{A collimated jet with a non-relativistic head ($\tilde L\ll1$) }
In this regime $\xi=1$ and the rest of the integration parameters
takes the following values: $\zeta=\varepsilon=\frac{5-\alpha}{3}$, $\varrho=\frac{3}{(3-\alpha)}$.
The solutions to the system's parameters is:
\begin{eqnarray}
z_h&=&\left(\frac{t^3L_j}{\rho\theta_0^4}\right)^{1/5}
\left[\frac{2^4}{3\pi}\varrho\zeta^2\right]^{1/5}\label{zh}\\
\beta_h&=&\left(\frac{L_j}{t^2\rho\theta_0^4}\right)^{1/5}
\left[\frac{2^4}{3\pi}\varrho\zeta^{-3}\right]^{1/5}
\frac{1}{c}\label{bh_nr}\\
P_c&=&\left(\frac{\rho^3L_j^2\theta_0^2}{t^4}\right)^{1/5}
\left[\frac{1}{6\pi}\varrho\zeta^{-3}\right]^{2/5}\\
r_c&=&z\theta_0\frac{1}{2\sqrt{\varrho}}\\
\beta_c&=&\beta_h\theta_0\frac{1}{2\sqrt{\varrho}}\\
\theta_c&=&\theta_0\frac{1}{2\sqrt{\varrho}}\label{th_c}\\
r_j&=&\left(\frac{t^4L_j^3\theta_0^8}{\rho^3}\right)^{1/10}
\left[\frac{2^4\pi\sqrt{\pi}}{3}\varrho\zeta^{-3}\right]^{-1/5}
\frac{1}{\sqrt{c}}\\
%
%
\theta_j&=&\left(\frac{L_j\theta_0^8}{t^2\rho}\right)^{1/10}
\left[\frac{2^8\sqrt\pi}{3}\varrho^2\zeta^{-1}\right]^{-1/5}
\frac{1}{\sqrt{c}}\\
\Gamma_1&=&\theta_0^{-1}\\
\hat z&=&\frac{2r_j}{\theta_0}=\sqrt{\frac{L_j}{\pi cP_c}}.
\end{eqnarray}

\subsection *{A collimated jet with a relativistic head ($1\ll\tilde L<\theta_0^{-4/3}$) }

In this regime $\zeta=1$, $\xi=\frac{5}{7-\alpha}$, $\varepsilon=\frac{5}{3+\alpha}$,
and $\varrho=\frac{3}{(3-\alpha)}$.
The solutions to the system's parameters is:

\begin{eqnarray}
z_h&\simeq&ct\\
\beta_h&\simeq&1\\
\Gamma_h&=&\left(\frac{L_j}{t^2\rho_a\theta_0^4}\right)^{1/10}
\left[\frac{1}{6\pi}\varrho\xi\varepsilon^{-2}\right]^{1/10}
\frac{1}{\sqrt{c}}\label{gh_r}\\
P_c&=&
\left(\frac{L_j^2\rho_a^3\theta_0^2}{t^4} \right)^ {1/5}
\left[\frac{1}{6\pi}\varrho\xi\varepsilon^{-2}\right]^{2/5}\\
r_c&=&\left(\frac{t^3L_j\theta_0}{\rho_a}\right)^{1/5}
\left[\frac{1}{6\pi}\varrho^{-3}\xi\varepsilon^3\right]^{1/5}\\
\beta_c&=&
\left(\frac{L_j\theta_0}{t^2\rho_a} \right)^ {1/5}
\left[\frac{1}{6\pi}\varrho^{-3}\xi\varepsilon^{-2}\right]^{1/5}\frac{1}{c}\\
\theta_c&=&\beta_c\\
r_j&=&\left(\frac{t^4L_j^3\theta_0^8}{\rho_a^3}\right)^{1/10}
\left[\frac{3\varepsilon^2}{16\pi^{3/2}\varrho\xi}\right]^{1/5}\frac{1}{\sqrt{c}}\\
\theta_j&=&\left(\frac{L_j^3\theta_0^8}{t^6\rho_a^3}\right)^{1/10}
\left[\frac{3\varepsilon^2}{16\pi^{3/2}\varrho\xi}\right]^{1/5}\frac{1}{c\sqrt{c}}\\
\Gamma_1&=&\theta_0^{-1}\\
\hat z&=&\frac{2r_j}{\theta_0}=\sqrt{\frac{L_j}{\pi cP_c}}.
\end{eqnarray}

The jet is collimated as long as $\hat z\leq z_h$. Substituting eqn. (B-12, B-15) in
eq. (B-22), and using the relation $\tilde L=\frac{4P_c}{\theta_0^2\rho_ac^2}$ (eq. \ref{tilde_L_strong}),
we get that the jet is collimated as long as
\begin{equation}
\tilde L\leq\theta_0^{-4/3}\left[\frac{16\pi^{3/2}\varrho\xi}{3\varepsilon^2}\right]^{2/3}.
\end{equation}

\subsection *{An un-collimated jet with a causally connected relativistic head ($\theta_0^{-4/3}\ll\tilde L\ll\theta_0^{-4}$)}

The integration parameters in this regime are the same as in the relativistic, collimated
regime, i.e. $\zeta=1$, $\xi=\frac{5}{7-\alpha}$, $\varepsilon=\frac{5}{3+\alpha}$
and , $\varrho=\frac{3}{(3-\alpha)}$.
But since the jet is now conical the solutions are:
\begin{eqnarray}
z_h&\simeq&ct\\
\beta_h&\simeq&1\\
\Gamma_h&=&\left(\frac{L_j\theta_0^2}{t^2\rho}\right)^{1/4}
\left(\frac{1}{4\pi c^5}\right)^{1/4}
\label{gh_r}\\
P_c&=&\left(\frac{L_j\rho_a^3\theta_0^2}{t^2}\right)^{1/4}
\left(\frac{\xi\varrho}{3\pi\zeta\varepsilon^2}\right)^{1/2}c^{3/4}\\
\beta_c&=&\left(\frac{L_j\theta_0^2}{t^2\rho_a}\right)^{1/8}
\left(\frac{\xi}{3\pi\zeta\varepsilon^2\varrho}\right)^{1/2}c^{-5/8}\\
r_c&=&\left(\frac{t^6L_j\theta_0^2}{\rho_a}\right)^{1/8}
\left(\frac{\xi\varepsilon^2}{3\pi\zeta\varrho}\right)^{1/2}c^{3/8}\\
\theta_c&=&\beta_c\\
\theta_j&=&\theta_0.
\end{eqnarray}

\subsection *{An un-collimated jet with an un-causal relativistic head ($4\theta_0^{-4}\ll\tilde L$)}

In this regime different parts in the head are not in causal connection and therefore
only a fraction $2/(\Gamma_h\theta_0)$ of the energy in the head goes into the cocoon.
The cocoon's pressure is relativistic and it pushes the edge of the cocoon to a velocity $\beta_c\rightarrow1$
which implies that our approximation of the cylindrical cocoon breaks down.
Moreover since the $\beta_c>c/\sqrt{3}$, the cocoon is no longer causally connected in the
lateral direction and the pressure is no longer uniform.
In this case we are able to provide the average energy density in the cocoon, by taking
the total energy that enters the cocoon and dividing it with $V_c$,
the cocoon's volume approximated as a sphere in this regime.
Under these approximations the model parameters are:
\begin{eqnarray}
z_h&\simeq&ct\\
\beta_h&\simeq&1\\
\Gamma_h&=&\left(\frac{L_j\theta_0^2}{t^2\rho}\right)^{1/4}
\left(\frac{1}{4\pi c^5}\right)^{1/4}
\label{gh_r}\\
\frac{E_c}{V_c}&=&\left(\frac{L_j\rho_a^3\theta_0^2}{t^2}\right)^{1/4}
\frac{3}{\sqrt{2}}c^{3/4}\\
\beta_c&=&1\\
r_c&=&ct\\
\theta_j&=&\theta_0.
\end{eqnarray}

When $\tilde L\gg4\Gamma_{j0}$ the Lorentz factor of the head is $\sim\Gamma_{j0}$,
and the reverse shock becomes Newtonian.
In this limit most of the energy that flows into the cocoon comes from the
material behind the forward shock which remains relativistic.
Energy enters the head through the forward shock at a rate:
$E_{ah}=\pi t^3\theta_0^2\Gamma_{j0}^2\rho_a^2c^5$,
and a fraction $2/(\Gamma_{j0}\theta_0)$ of that flows into the cocoon.
Approximating the cocoon's volume as a sphere we get:
\begin{eqnarray}
\Gamma_h&=&\Gamma_{j0}\\
\frac{E_c}{V_c}&=&\frac{3}{2}\Gamma_{j0}\theta_0\rho_ac^2,
\end{eqnarray}
where all the rest of the parameters remains as before.
Note that in this regime the average energy density in the cocoon remains constant.


\begin{thebibliography}{}

\baselineskip=15.8pt


\bibitem[\protect\citeauthoryear{Aloy et al.}{1999}]{Aloy99}
Aloy M.~A., Ib{\'a}{\~n}ez J.~M.~\^{}., Mart{\'{\i}} J.~M.~\^{}., G{\'o}mez
J.-L., M{\"u}ller E., 1999, ApJ, 523, L125
\bibitem[\protect\citeauthoryear{Band et al.}{1993}]{Band93} Band D., et al.,
1993, ApJ, 413, 281
\bibitem[\protect\citeauthoryear{Begelman \& Cioffi}{1989}]{BegelCiof89}
Begelman M.~C., Cioffi D.~F., 1989, ApJ, 345, L21
\bibitem[\protect\citeauthoryear{Blandford \& Rees}{1974}]{BlandRees74}
Blandford R.~D., Rees M.~J., 1974, MNRAS, 169, 395
\bibitem[\protect\citeauthoryear{Bromberg \& Levinson}{2007}]{BL07} Bromberg O.,
\& Levinson A., 2007, ApJ, 671, 678; BL07.
\bibitem[\protect\citeauthoryear{Bromberg \& Levinson}{2009}]{BL09} Bromberg O.,
Levinson A., 2009, ApJ, 699, 1274; BL09.
\bibitem[\protect\citeauthoryear{Bromberg et al.}{2011}]{Bromberg11} Bromberg O.,
Nakar U., Piran T., Sari R., 2011, in prep.
\bibitem[\protect\citeauthoryear{Bulbul et al.}{2010}]{Bulbul10}
Bulbul G.~E., Hasler N., Bonamente M., Joy M., 2010, ApJ, 720, 1038
\bibitem[\protect\citeauthoryear{Capelo, Natarajan,\& Coppi}{2010}]
{Capelo10} Capelo P.~R., Natarajan P., Coppi P.~S., 2010, MNRAS, 407, 1148
\bibitem[\protect\citeauthoryear{Crowther}{2007}]{Crowther07}
Crowther P.~A., 2007, ARA\&A, 45, 177
\bibitem[\protect\citeauthoryear{Eichler}{1994}]{Eichler94} Eichler, D. 1994,
Astrophys. J. Supp., 90, 877
\bibitem[\protect\citeauthoryear{Eichler \& Levinson}{2000}]{EL00} Eichler, D.,
\& Levinson, A. 2000, ApJ, 529, 146
\bibitem[\protect\citeauthoryear{Falle}{1991}]{Falle91} Falle
S.~A.~E.~G., 1991, MNRAS, 250, 581
\bibitem[\protect\citeauthoryear{Granot et al.}{2000}]{Granot00}
Granot J., Miller M., Piran T., Suen W.-M., 2000, AIPC, 526, 540
\bibitem[\protect\citeauthoryear{Hughes, Miller, \& Duncan}{2002}]{Hughes02}
Hughes P.~A., Miller M.~A., Duncan G.~C., 2002, ApJ, 572, 713
\bibitem[\protect\citeauthoryear{Kaiser\& Alexander}{1997}]{KaisAlex97}
Kaiser C.~R., Alexander P., 1997, MNRAS, 286, 215
\bibitem[\protect\citeauthoryear{Komissarov \& Falle}{1997}]{KomisFall97}
Komissarov S.~S., Falle S.~A.~E.~G., 1997, MNRAS, 288, 833
\bibitem[\protect\citeauthoryear{Lazzati \& Begelman}{2005}]{LazzBeg05} Lazzati D.,
\& Begelman M., 2005, ApJ, 629, 903,
\bibitem[\protect\citeauthoryear{Levinson \& Bromberg}{2008}]{LB08} Levinson A.,
\& Bromberg O., 2008, PhRvL, 100, 131101
\bibitem[\protect\citeauthoryear{MacFadyen \& Woosley}{1999}]{MacWoos99}
MacFadyen A.~I., Woosley S.~E., 1999, ApJ, 524, 262
\bibitem[\protect\citeauthoryear{Matzner}{2003}]{Matzner03} Matzner C.~D.,
2003, MNRAS, 345, 575
\bibitem[\protect\citeauthoryear{Matzner \& McKee}{1999}]{MatzMck99}
Matzner C.~D., McKee C.~F., 1999, ApJ, 510, 379
\bibitem[\protect\citeauthoryear{Marti et al.}{1995}]{Marti95}
Marti J.~M.~A., Muller E., Font J.~A., Ibanez J.~M., 1995, ApJ, 448, L105
\bibitem[\protect\citeauthoryear{Marti et al.}{1997}]{Marti97}
Marti J.~M.~A., Mueller E., Font J.~A., Ibanez J.~M.~A., Marquina A., 1997,
ApJ, 479, 151
\bibitem[\protect\citeauthoryear{M{\'e}sz{\'a}ros et al.}{1993}]{Meszaros93}
M{\'e}sz{\'a}ros P., Laguna P., Rees M.~J., 1993, ApJ, 415, 181
\bibitem[\protect\citeauthoryear{M{\'e}sz{\'a}ros \& Rees}{2000}]{Meszaros00}
M{\'e}sz{\'a}ros P., \& Rees M.~J., 2000, ApJ, 530, 292
\bibitem[\protect\citeauthoryear{M{\'e}sz{\'a}ros \& Waxman}{2001}]{MW01}
M{\'e}sz{\'a}ros P., \& Waxman E., Phys. Rev. Lett. {\bf 87}, 171102 (2001)
\bibitem[\protect\citeauthoryear{Miller-Jones, Fender \& Nakar}{2006}]{Miller06}
Miller-Jones J.~C.~A., Fender R.~P., Nakar E., 2006, MNRAS, 367, 1432
\bibitem[\protect\citeauthoryear{Mizuta \& Aloy}{2009}]{Mizuta09}
Mizuta A., Aloy M.~A., 2009, ApJ, 699, 1261
\bibitem[\protect\citeauthoryear{Morsony, Lazzati,\& Begelman}{2007}]{Morsony07}
Morsony B.~J., Lazzati D., Begelman M.~C., 2007, ApJ, 665, 569
\bibitem[\protect\citeauthoryear{Nakar et al.}{2005}]{Nakar05} Nakar E., Piran
T., \& Sari R., 2005, ApJ, 635, 516
\bibitem[\protect\citeauthoryear{Paczynski}{1990}]{Paczynski90} Paczynski B.,
1990, ApJ, 363, 218
\bibitem[\protect\citeauthoryear{Piran}{2000}]{Piran00} Piran
T., 2000, PhR, 333, 529
\bibitem[\protect\citeauthoryear{Piran, Shemi, \& Narayan}{1993}]{Piran93}
Piran T., Shemi A., Narayan R., 1993, MNRAS, 263, 861
\bibitem[\protect\citeauthoryear{Sari\& Piran}{1995}]{Sari95} Sari R., Piran T., 1995, ApJ, 455, L143
\bibitem[\protect\citeauthoryear{Scheuer}{1974}]{Scheuer74}
Scheuer P.~A.~G., 1974, MNRAS, 166, 513
\bibitem[\protect\citeauthoryear{Shemi \& Piran}{1990}]{ShemiPiran90} Shemi A.,
Piran T., 1990, ApJ, 365, L55
\bibitem[\protect\citeauthoryear{Woosley}{1993}]{Woosley93}
Woosley S.~E., 1993, ApJ, 405, 273
\bibitem[\protect\citeauthoryear{Woosley \& Heger}{2006}]{Woosley06} Woosley S.~E., Heger A., 2006, ApJ, 637, 914
\bibitem[\protect\citeauthoryear{Zhang, Woosley, \& MacFadyen}{2003}]{Zhang03}
Zhang W., Woosley S.~E., MacFadyen A.~I., 2003, ApJ, 586, 356
\end{thebibliography}
\end{document}